# High-Q Nanophotonic Resonators on Diamond Membranes using Templated Atomic Layer Deposition of TiO$_2$

*Amy Butcher[1], Xinghan Guo[1], Robert Shreiner[2], Nazar Delegan[3], Kai Hao[1], Peter J. Duda III[1], David D. Awschalom[1,3], F. Joseph Heremans[1,3], Alexander A. High[1,3]\**

[1] Pritzker School of Molecular Engineering, University of Chicago, Chicago, Illinois 60637, USA

[2] Department of Physics, University of Chicago, Chicago, Illinois 60637, USA

[3] Center for Molecular Engineering and Materials Science Division, Argonne National Laboratory, Lemont, Illinois 60439, USA

\* To whom correspondence should be addressed: ahigh@uchicago.edu

TABLE OF CONTENTS GRAPHIC:

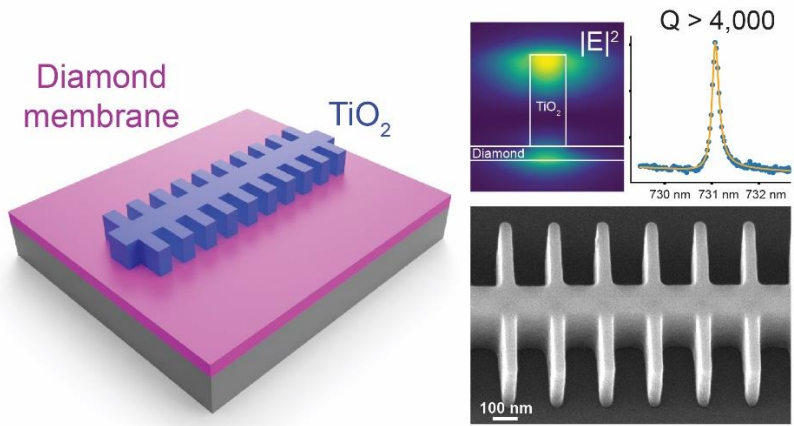

ABSTRACT: Integrating solid-state quantum emitters with nanophotonic resonators is essential for efficient spin-photon interfacing and optical networking applications. While diamond color centers have proven to be excellent candidates for emerging quantum technologies, their integration with optical resonators remains challenging. Conventional approaches based on etching resonators into diamond often negatively impact color center performance and offer low device yield. Here, we developed an integrated photonics platform based on templated atomic layer deposition of TiO$_2$ on diamond membranes. Our fabrication method yields high-performance nanophotonic devices while avoiding etching wavelength-scale features into diamond. Moreover, this technique generates highly reproducible optical resonances and can be iterated on individual diamond samples, a unique processing advantage. Our approach is suitable for a broad range of both wavelengths and substrates and can enable high-cooperativity interfacing between cavity photons and coherent defects in diamond or silicon carbide, rare earth ions, or other material systems.

MAIN TEXT: Diamond color centers are a promising basis for emerging quantum technologies due to their robust spin coherence, optical addressability, and on-chip integration capability[1]. Nanophotonic resonators are critical for building quantum networks based on these centers[2], as they enable deterministic interactions between individual optical photons and electronic spin states. By providing a light-matter interface, integrated nanophotonic devices can mediate entanglement between separated color centers[3] while simultaneously enhancing their quantum emission rates and improving photon indistinguishability through the Purcell effect[4]. In order to capitalize on the potential of integrated photonics for quantum networking, a scalable and deterministic method of device fabrication is needed.

Current methods for integrating diamond color centers with nanophotonic resonators[5] suffer from low yield and are often incompatible with the fabrication of other on-chip devices. These approaches rely on 1) angled or quasi-isotropic etching into bulk diamond to generate suspended structures[6-9], 2) etching devices vertically into sub-wavelength thick diamond membranes[10-12], or 3) etching devices vertically into gallium phosphide-diamond heterostructures[13-15]. In the last decade, these techniques have enabled landmark demonstrations of zero-phonon line (ZPL) enhancement[11,13,16] and coherent light-matter interactions in single devices[17]. However, carving wavelength-scale features into diamond damages and strains the crystal lattice and brings poorly controlled surface states into close proximity with color centers[18-20]. These effects significantly degrade color center spin coherence compared to bulk values[16] and consequently limit operational device yield. Additionally, suspended structures generated from bulk diamond are subject to geometrical constraints which complicate their integration with on-chip single photon detectors, electronics, or other devices that could enhance quantum network functionality.

Scattering from surface roughness creates further challenges for diamond nanophotonics, particularly in the visible wavelength regime. Along with the nitrogen-vacancy center[21], with its ZPL at 637 nm, several group-IV color centers operate in the visible, including the silicon-vacancy (ZPL 737 nm)[22,23], germanium-vacancy (ZPL 602 nm)[24,25], and tin-vacancy (ZPL 620 nm)[26] centers. For a given root-mean-square roughness σ, scattering loss[27] is proportional to $\sigma^2/\lambda^3$. Accordingly, to overcome this loss at short wavelengths, optical devices designed to interact with diamond color centers ought to be exceptionally smooth. Because diamond is both extremely hard and inert, it is difficult to micromachine its surfaces smoothly, and quality factors for diamond nanophotonics are often limited by surface scattering as a result.

An alternative approach is to build nanophotonic resonators on top of diamond membranes rather than out of diamond itself. This strategy avoids micromachining diamond and consequently minimizes crystal distortion and sidewall roughness in final structures. Importantly, with sufficiently thin membranes, overlying devices retain the potential to enable coherent single photon-spin interactions. With this application in mind, we developed a nanofabrication platform based on templated atomic layer deposition (ALD)[28-30] of $TiO_2$ which yields smooth, high-Q optical resonators without a substrate or sidewall etching step. We chose $TiO_2$ as the guiding material for our devices because it has a high refractive index (n > 2.3) over a broad range of wavelengths and its relatively large bandgap (~3.3 eV) makes it suitable for visible device operation[28,31,32].

In this Letter, we detail our fabrication process and provide examples of nanophotonic devices, i.e. ring resonators and 1D photonic crystal cavities, that were created using this technique. These results represent the first application of templated ALD to on-chip waveguiding optics. With uniquely low optical loss and excellent processing flexibility, this technique may prove beneficial for a range of integrated photonics applications, both classical and quantum. To demonstrate our platform's potential to enable spin-photon interfacing with quantum defects, we fabricated high-Q resonators on a 50 nm-thick single crystal diamond membrane. With cavity modes concentrated largely within the membrane itself, we show this platform can enable large defect-cavity cooperativities for efficient spin-photon interfacing. Notably, this fabrication procedure can be iterated on a single diamond sample without substrate damage, facilitating precise alignment of cavities to color centers.

We begin our fabrication process (illustrated in Fig. 1a-c) by patterning device templates into poly(methyl methacrylate) (PMMA) via electron beam lithography. Before filling these patterns with $TiO_2$, we use a brief, three-second oxygen plasma exposure to remove residual polymer adsorbates at the bottom of the templates which can increase loss in final devices. In contrast to prolonged plasma etching, this step does not damage the underlying substrate and simply ensures the template is clean and ready to be filled.

Next, we use ALD to conformally fill the resist templates with high-index $TiO_2$[33]. To avoid reflowing the resist, the deposition occurs at 90 C, which is below the glass transition temperature of most resists (~105 C for PMMA). A low chamber temperature also ensures that the $TiO_2$ remains amorphous rather than rutile or anatase. These other phases of $TiO_2$ develop during ALD at higher deposition temperatures and form grain boundaries, which can lead to scattering loss[32]. We significantly over-fill the device templates to help planarize the top $TiO_2$ surface, where a slight crease forms (Fig. 1d) due to the conformal filling profile.

After deposition, we etch away the excess TiO$_2$ with inductively coupled plasma reactive ion etching (ICP RIE) to expose the resist underneath. Importantly, because we do not etch through the resist template during this step, neither the substrate nor the device sidewalls are etched by the ICP. After removing the TiO$_2$ overfill, we chemically strip the remaining resist and any etch residues (Nanostrip, MicroChem) to reveal the templated devices (Fig. 1e-f). Lastly, we anneal the TiO$_2$ structures on a hot plate at 250 C for two hours, which we found was critical for reducing material optical loss[33].

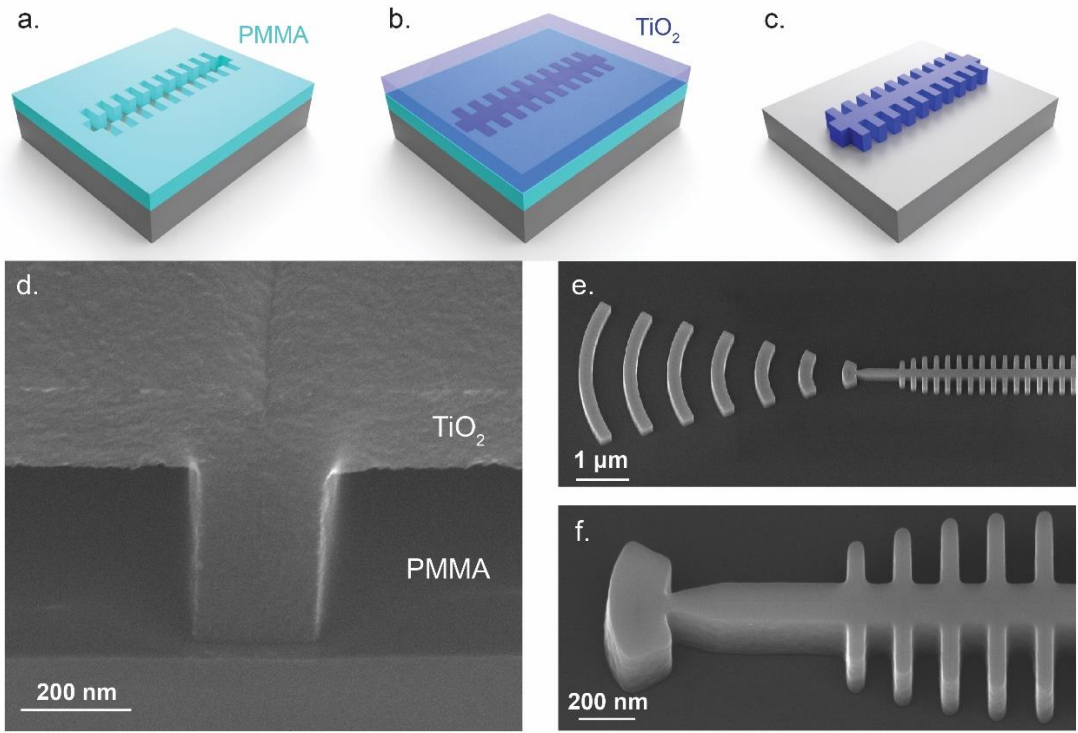

**Figure 1.** Nanofabrication process. (a) Nanophotonic device templates are patterned into e-beam resist and then (b) conformally filled with ALD TiO$_2$. (c) Overfilled TiO$_2$ is etched back and resist template is stripped. (d) 45 degree cross-sectional scanning electron microscope (SEM) image of a resist trench over-filled with TiO$_2$. (e, f) 30 degree SEM images of completed nanophotonic structures.

To demonstrate the performance of this fabrication platform, we first built high-Q microring resonators on fused silica substrates. Ring resonators can function as spectral filters in integrated photonic networks[34] and are attractive for applications in sensing[35] and chiral quantum optics[36]. We fabricated resonators in an add-drop geometry, where two linear bus waveguides couple to either side of a 5um-radius circular loop (Fig. 2a and d). The waveguides were 300 nm wide and 250 nm tall, with coupling distances of 375 nm between the buses and resonator (Fig. 2b). At each end of the bus waveguides, we designed and fabricated grating couplers that can couple up to ~48% of incident light into and out of the waveguides[33].

To probe the ring resonator spectra, we coupled a broadband excitation source into one bus waveguide and measured transmitted signal through the drop port. As depicted in the inset of Fig. 2a, a bright outcoupling spot appears at the opposite end of the excitation waveguide, where most of the broadband source is transmitted, while a weaker spot containing only resonant frequencies is detected at the drop port. For these 5 um-radius devices, the spectra feature standing wave resonances separated by approximately 5.9 nm. Because the resonance linewidths are too narrow to be fully resolved by our spectrometer, we scan a tunable laser across each peak and measure transmitted intensity to accurately quantify quality factors for these devices. Typical Q values were 20,000-30,000. While it is challenging to compare resonators with different radii and operating wavelengths, these values are competitive with state-of-the-art quality factors reported elsewhere[11,31,35]. This result demonstrates that our low-loss templated ALD platform can yield high performance optical resonators.

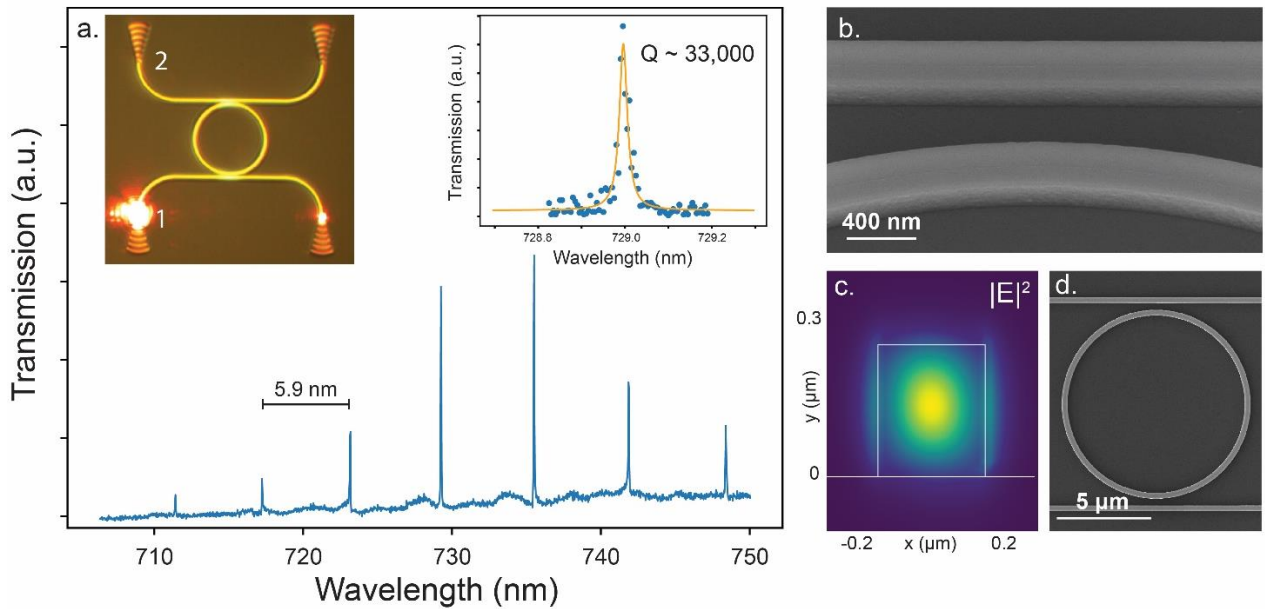

**Figure 2.** Microring resonators on fused silica. (a) Transmission spectrum for a ring resonator with 5 μm radius. Left inset: camera view of laser in- and out-coupling. Excitation and collection ports are labelled one and two, respectively. Right inset: high-resolution resonance scan showing Q = 33,260. (b) 30 degree tilted SEM image of waveguide coupling region for device in (a). (c) Simulated cross-sectional electric field intensity of the TE resonator mode. (d) Top-down SEM image of ring resonator in (a), (b).

We next designed and fabricated 1D photonic crystal cavities, which have significantly lower mode volumes than ring resonators and can consequently induce larger Purcell enhancement for a given quality factor. Our design features a 150 nm-wide and 250 nm-tall central waveguide with fins that periodically modulate the effective refractive index along its length (Fig. 3a). The cavity region is formed at the center of the waveguide, where the pitch of the fins gradually decreases by 10%. Quality factors for these structures depend on the number of fins on either side of the central cavity region, with more fins corresponding to higher reflectivity and higher Qs. Unlike the ring resonators, in which the resonator modes are concentrated at the center of the waveguide cross-section (Fig. 2c), we design the 1D cavities to accommodate a single mode concentrated near the waveguide's top and bottom (Fig. 3b-c). Notably, this design increases the electric field intensity near underlying elements, which enhances light-matter coupling in those regions.

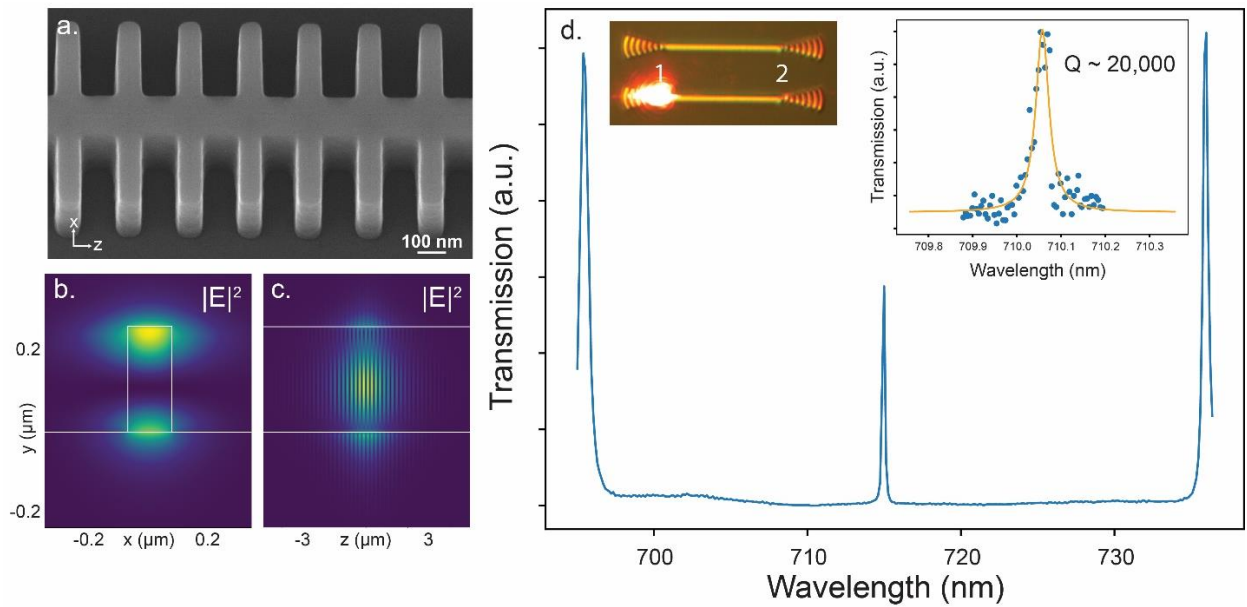

**Figure 3.** 1D cavities on fused silica. (a) 30 degree tilted SEM image of 1D photonic crystal cavity. (b), (c) Simulated cross-sectional electric field intensity of the cavity mode across the cavity center and along the cavity length, respectively. (d) Transmission spectrum for a 1D cavity on fused silica with 20-fin reflectors. Left inset: camera view of laser in- and out-coupling. Excitation and collection ports are labelled one and two, respectively. Right inset: high-resolution resonance scan for 25-fin device in (a) showing Q = 19,640.

To validate the performance and reproducibility of these devices, we first characterized their spectra on fused silica substrates. Table 1 shows that cavity resonance wavelengths are highly deterministic, with standard deviations in the range of 0.5-1.4 nm. These statistics were averaged over ten devices with 25-fin reflectors at each of three wavelengths. Although it was not possible to scan our tunable laser at wavelengths lower than 700 nm, spectrometer data gives a lower bound for quality factors, which exceed 6,000 for even the shortest-wavelength resonances. By scanning over device resonances near 710 nm, we measured quality factors approaching 20,000, more than twice what is approximated by spectrometer-limited data (Fig 3d). These results, combined with SEM images, indicate that device sidewalls are sufficiently smooth for high-Q visible frequency operation.

**Table 1.** Resonance statistics averaged over ten 25-fin 1D photonic crystal cavities at each of three wavelengths.

| Mean $\lambda$ (nm) | Std. dev. $\lambda$ (nm) | Mean Q[a] |
|---|---|---|
| 606.499 | 0.623 | 6,040 |
| 655.804 | 0.561 | 7,210 |
| 710.341 | 1.326 | 9,070 |

[a] Due to limited spectrometer resolution, quality factors are interpreted as lower bounds only

As a demonstration of this platform's potential to interface with color centers in diamond, we fabricated 1D cavities on a 50 nm-thick single crystal diamond membrane (Fig. 4a-b). The membrane was generated via ion bombardment[37] and chemical vapor deposition overgrowth. It was then adhered to a Si carrier chip using hydrogen silsesquioxane (HSQ)[38,39], which we subsequently annealed to produce a low index substrate (see Methods). Our fabrication procedure for the resonators on the membrane was identical to the fused silica case, except we built devices slightly taller (300 nm instead of 250 nm) to maintain waveguiding on the higher-index (n = 2.4) diamond.

For 1D cavities on the diamond membrane, we measured quality factors reaching 4,400 (Fig. 4c), with the majority of devices showing Q ~ 4,000[33]. Possible reasons for the discrepancy in quality factors between these cavities and those on silica substrates include scattering loss within the HSQ layer, absorption loss at the $TiO_2$-diamond interface, and absorption loss within the diamond membrane itself[33]. The experimentally measured quality factors are sufficient for high-efficiency spin-photon interfacing with diamond color centers. Simulated cavities operating at the silicon-vacancy center's ZPL (737 nm) exhibit mode volumes of $2.0(\lambda/n)^3$. Combining this value with Q = 4,400, we calculated Purcell factors[33] of up to 175 at the cavity mode maximum and 115 within the underlying diamond, where a large fraction of the cavity mode intensity is concentrated (Fig. 4d). In terms of cooperativity, we estimate values in the range 1-10 based on comparisons with previous work[33,40,41].

Finally, we note that this fabrication process can be iterated multiple times on one membrane and can generate a large number of resonators in arbitrary orientations. After $TiO_2$ resonators are fabricated, they can be mechanically removed via ultrasonication without removing or damaging the underlying membrane. Therefore, if alignment to a particular color center is not ideal, or if a fabrication step fails, additional attempts can be made. The data displayed in Fig. 4c were acquired from the third round of cavity fabrication on a single membrane sample. Each of these three rounds yielded devices with typical quality factors Q ~ 4,000. Meanwhile, atomic force microscopy scans do not show significant difference to the surface roughness of the diamond between fabrication rounds. This flexibility is a distinct advantage compared to diamond etching methods, which require re-polishing and potentially re-implanting the diamond to generate new color centers for each iteration. Moreover, as pictured in Fig. 4e, many devices can be fabricated at once and in arbitrary positions and orientations relative to one another, which is beneficial for on-chip device multiplexing.

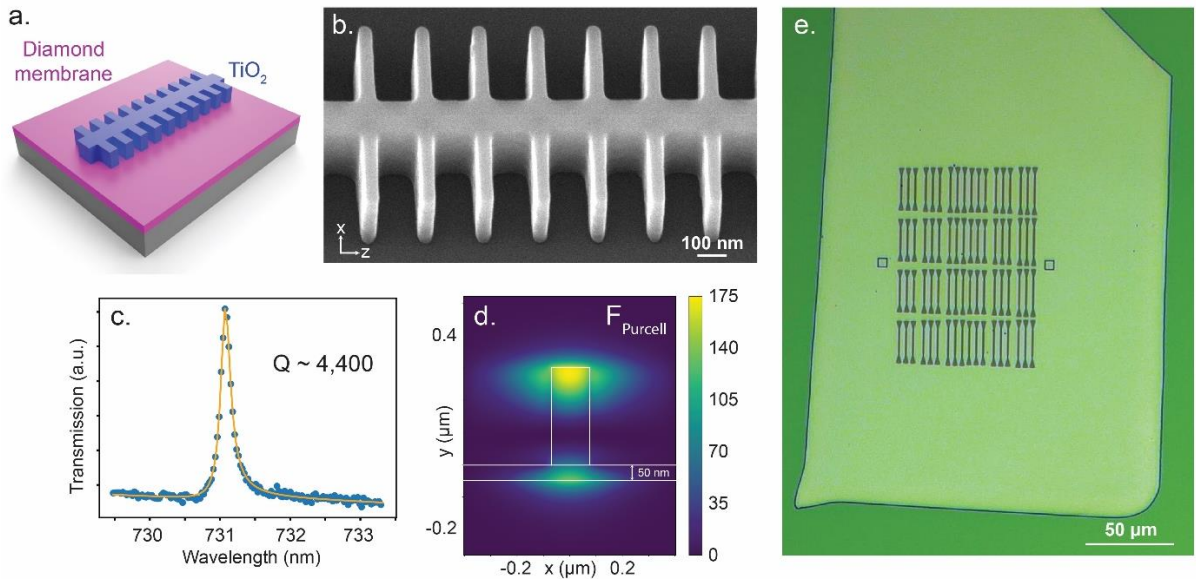

**Figure 4.** 1D cavities on diamond membranes. (a) Schematic of a 1D photonic crystal cavity on a diamond membrane. (b) 30 degree tilted SEM image of a cavity on a 50 nm-thick diamond membrane. (c) Transmission spectrum of device pictured in (b) showing Q = 4,410. (d) Simulated maximum Purcell

factor ($F_{Purcell}$) across the cavity center given a quality factor of 4,400. (e) Microscope image of 72 cavities fabricated on a 50nm-thick diamond membrane with 200 μm x 300 μm lateral area.

Integration of $TiO_2$ nanophotonic devices with diamond membranes offers a robust and non-invasive approach to optically addressing color centers that is more scalable and deterministic than conventional etched nanophotonics. Our ALD-based fabrication platform offers the dual benefits of minimized surface roughness and highly reproducible resonator spectra. Looking forward, we aim to use this platform to build quantum networks by aligning cavities to individual color centers within membranes and gas tuning their resonances to the defects' zero phonon line[42]. With high quality factors and low mode volumes, we anticipate that efficient spin-photon interfacing will be achieved. Finally, we note that this platform is flexible to both wavelength and substrate, allowing for integration with defects in silicon carbide[43-45], rare earth ions[46], or other material systems beyond diamond

METHODS:

**Electron beam lithography:** E-beam resist (950K PMMA A4, MicroChem) was spin-coated to a thickness of 270 nm-330 nm depending on application. Next, a conduction layer of 20 nm Au was thermally evaporated on top to avoid charge build-up during lithography. To further mitigate charging effects, we used a relatively low beam current (0.3 nA), multipass exposure (GenISys BEAMER), and a writing order which reduced the time the electron beam dwelled continuously in one region. Resist was exposed with a total dose of 1200 uC/cm2 at 100 kV (Raith EBPG5000 plus). After exposure, we removed the conduction layer with TFA gold etchant (Transene) and developed the resist in a 1:3 MIBK:IPA solution on a cold plate at 7 C for 90 s, followed by a 60 s IPA stopper and DI water rinse.

**Atomic layer deposition:** Before ALD, we used a three-second plasma exposure in an ICP RIE tool (PlasmaTherm Apex) with 10 sccm O2 and 50 W ICP power to remove residual polymer. With this recipe, the PMMA etch rate is roughly 2.5 nm/s. For the $TiO_2$ deposition, we used a commercial thermal ALD chamber (Veeco/Cambridge Savannah ALD). Amorphous $TiO_2$ was deposited at 90 C using tetrakis(dimethylamido) titanium (TDMAT) and water, alternately pulsed for 0.08 s and 0.10 s, respectively. 100 sccm of $N_2$ flowed continuously during deposition and wait times between precursor pulses were 8 s long. Deposition rates were typically 0.6 A/cycle.

**ICP etch procedure:** We removed over-filled $TiO_2$ through chlorine-based ICP RIE etching (PlasmaTherm Apex) with 150 W substrate bias, 400 W ICP power, 12 sccm $Cl_2$ and 8 sccm BCl. Etch rates were typically 1.5-1.7 nm/s.

**SEM imaging:** Following thermal deposition of a 5 nm Cr conduction layer, nanophotonic structures were imaged using a Carl Zeiss Merlin FE-SEM.

**FDTD simulations:** Ring resonators, photonic crystal cavities, and grating couplers were simulated using Lumerical finite difference time domain software.

**Transmission spectroscopy:** We used a home-built confocal microscopy set-up with separate collection and excitation channels to perform transmission spectroscopy. A pulsed supercontinuum source (430-2400 nm, SC-OEM YSL Photonics) and spectrometer (1200 g/mm, Princeton Instruments) were used for broadband measurements. To generate high-resolution scans over individual cavity resonances, we swept a 50 kHz-linewidth, tunable CW laser (MSquared) for excitation and used an avalanche photodiode (Excelitas) for detection.

**Diamond membranes:** 500 nm-thick diamond membranes were generated through ion bombardment[34] and were overgrown with chemical vapor deposition at Argonne National Lab. After electrochemical etching of the ion-damaged layer, suspended membranes were removed and flipped with a PDMS stamp. They were then adhered to a Si carrier using ~500 nm of HSQ resist and annealed for 8 hours at 420 C in Ar gas. Finally, the membranes were back-etched to the desired thickness using ICP etching with 25 sccm Ar, 40 sccm $Cl_2$, 400 W ICP power and 250 W bias power. Etch rates were typically 1.2-1.4nm/s.

ASSOCIATED CONTENT:

**Supporting Information.** Additional characterization of as-deposited and annealed $TiO_2$ films, design and simulation of grating couplers, discussion of optical loss on diamond membrane-HSQ heterostructures, Purcell factor calculation, and cooperativity estimate details. Supporting information is available in pdf format free of charge via pubs.acs.org.

ACKNOWLEDGEMENTS:

This work made use of the Pritzker Nanofabrication Facility of the Pritzker School of Molecular Engineering at the University of Chicago, which receives support from Soft and Hybrid Nanotechnology Experimental (SHyNE) Resource (NSF ECCS-1542205), a node of the National Science Foundation's National Nanotechnology Coordinated Infrastructure. This work also made use of shared facilities supported by the NSF MRSEC Program under Grant No. DMR-0820054.

Diamond CVD growth at Argonne National Lab is supported by the US Department of Energy, Office of Science, Basic Energy Sciences, Materials Sciences and Engineering Division. The materials characterization was supported as part of the Center for Novel Pathways to Quantum Coherence in Materials, an Energy Frontier Research Center funded by the U.S. Department of Energy, Office of Science, Basic Energy Sciences. Additional funding was provided by DARPA D18AC00015KK1932, AFOSR FA9550-19-1-0358, and the Boeing company. FDTD simulations were completed with resources provided by the University of Chicago's Research Computing Center.

A.B. acknowledges support from the NSF Graduate Research Fellowship under Grant No. DGE-1746045. The authors thank R. Devlin, M. Dolejsi, and P. Bennington for helpful discussions.